%% file: main.tex
\documentclass[runningheads]{llncs}
\usepackage{xcolor}
\usepackage{amsmath}
\usepackage{enumitem}
\usepackage{fancyvrb}
\usepackage{alltt}
\usepackage{tikz}
\usetikzlibrary{shapes.geometric, arrows}
\usetikzlibrary{positioning}
\usetikzlibrary{backgrounds}
\usepackage{algorithm}
\usepackage{algpseudocode}
\usepackage{algcompatible}
\usepackage{listings, xcolor}
\usepackage{pgfplots, pgfplotstable}
\pgfplotsset{compat=1.8}
\usepackage{booktabs}
\usepackage{forest}
\usepackage{siunitx}
\usepackage{changepage}
\usepackage{comment}
\usepackage{subcaption}
\usepackage{rotating}

\algtext*{ENDWHILE}
\algtext*{ENDFOR}

\newcommand\logicTurnstile{\mathop{\vcenter{\hbox{$:$}}-}}

\definecolor{dkgreen}{rgb}{0,0.6,0}
\definecolor{gray}{rgb}{0.5,0.5,0.5}
\definecolor{mauve}{rgb}{0.58,0,0.82}
\lstdefinestyle {DatalogStyle} {frame=tb,
  language=C++,
  aboveskip=3mm,
  belowskip=3mm,
  escapechar=`,
  showstringspaces=false,
  columns=flexible,
  basicstyle={\small\ttfamily},
  numbers=left,
  numberstyle=\tiny\color{gray},
  keywordstyle=\color{blue},
  commentstyle=\color{dkgreen},
  stringstyle=\color{mauve},
  breaklines=true,
  breakatwhitespace=true,
  tabsize=3
}

\begin{document}
\title{The Choice Construct in the Souffl\'e Language}
\titlerunning{The Choice Construct in the Souffl\'e Language}
\author{Xiaowen Hu\inst{1,2}\orcidID{0000-0002-4577-3360} \and
Joshua Karp\inst{1,3}\orcidID{0000-0002-4704-2182} \and
David Zhao\inst{1,4}\orcidID{0000-0002-3857-5016} \and
Abdul Zreika\inst{1,5}\orcidID{0000-0001-8812-5067} \and
Xi Wu\inst{1,6}\orcidID{0000-0001-5795-9798} \and
Bernhard Scholz\inst{1,7}\orcidID{0000-0002-7672-7359}}

\authorrunning{X. Hu et al.}

\institute{The University of Sydney, Australia \and
\email{xihu5895@uni.sydney.edu.au} \and
\email{jkar4969@uni.sydney.edu.au} \and
\email{dzha3983@uni.sydney.edu.au} \and
\email{azre6702@uni.sydney.edu.au} \and
\email{xi.wu@sydney.edu.au} \and
\email{bernhard.scholz@sydney.edu.au}}
\maketitle              

\begin{abstract}
Datalog has become a popular implementation language for 
solving large-scale, real world problems, including bug finders,  network analysis tools,  and  disassemblers. These applications express
complex behaviour with hundreds of relations and rules that often
require a non-deterministic choice for tuples in relations to express worklist algorithms.

This work is an experience report that describes 
the implementation of a \textit{choice} 
construct in the Datalog engine Souffl\'e.
With the choice construct we can 
express worklist algorithms such as spanning trees in a few lines of code.
We highlight the differences between rule-based choice as described in prior work, and relation-based choice introduced by this work. We show that a choice construct enables certain worklist algorithms to be computed up to 10k$\times$ faster than having no choice construct.  
\end{abstract}

\section{Introduction}

Datalog and other logic specification languages~\cite{jordan2016souffle,flix16,logicblox15,z311} have become popular in recent years for implementing bug finders, static program analysis frameworks~\cite{jordan2016souffle,pointsto15}, network analysis tools~\cite{eq10,de11}, security analysis tools~\cite{mv05}
and business applications~\cite{logicblox15}. 
For these applications, logic programming 
is used as a domain specific language 
to allow programmers to express complex program behavior 
succinctly, while enabling rapid-prototyping for scientific and
industrial applications in a declarative fashion. 
For example, logic programming has gained traction in the
area of program analysis due to its flexibility in building custom
program analyzers~\cite{jordan2016souffle}, points-to  
analyses for Java programs~\cite{bravenboer2009strictly}, and security analysis
for smart contracts~\cite{mm18,icse19}.

Although modern Datalog implementations such as Souffl\'e~\cite{scholz2016souffle} have constructs (e.g., functors) that make Datalog Turing-equivalent, certain 
classes of algorithms are hard to implement. 
For example, worklist algorithms~\cite{worklist02} that 
are commonly found in compilers and productivity tools~\cite{dragonbook}, 
are challenging since they require a 
non-deterministic choice from a set. 
Without the notion of choice, programmers must manually 
introduce an (arbitrary) ordering on a set and select the elements inductively to simulate this choice. The ordering and the inductive selection in Datalog requires dozens of rules and can be highly inefficient. 

In database literature~\cite{initialchoice,choice,choiceaggregates,textbook,expressivepower}, there have been Datalog extensions for non-deterministic choice.
In the work of Krishnamurthy, 
Naqvi, Greco and Zaniolo, the non-determinism 
is enforced operationally by introducing 
functional dependency constraints on 
relations.
A functional dependency constraint enforces that a particular subset of values in each tuple (the key) can only occur once in the relation. For example, an ternary relation $(x, y, z)$ with the functional dependency constraint $(x, y) \rightarrow z$ ensures that the two tuples $(1, 2, 3)$ and $(1, 2, 4)$ cannot simultaneously exist in the relation, since they both contain the same values $(1, 2)$ for the key $(x, y)$. In this system, any tuple in the relation causes all subsequent tuples that violate the functional dependency constraint to be rejected from being inserted into the relation.

In this work, we report on the experience of implementing a
choice construct in Souffl\'e~\cite{scholz2016souffle,jordan2016souffle}
and show (1) the simplicity of its semantics, 
(2) its ease of implementation,  and (3) its efficiency in contrast to having no choice construct in the language. 
Prior work on choice has introduced functional 
dependencies as local,  rule-based constraints, where the 
permissible tuples of a relation  are only constrained on a 
rule-by-rule basis~\cite{choice}. That work
must be seen in the context of database research in the 90s
that typically have a small number of rules.
Souffl\'e programs have different 
characteristics, consisting of hundreds of rules and relations~\cite{bravenboer2009strictly}, where the
relations are held in memory.
For such applications, a rule-based choice
becomes tedious and error prone because the 
functional dependency constraint may need to be repeated per rule.
Hence, we introduce a new variant of choice called
\emph{relation-based choice}. 
A relation-based choice makes the
underlying auxiliary relations of a ruled-based choice~\cite{Giannotti91} explicit to the programmer.
This approach is  more
amenable for logic programming with many
relations/rules to ease the burden for the programmer.

The contributions of our paper are summarized as follows:

\begin{itemize}
    \item We introduce a relation-based choice construct for the Souffl\'e (a Datalog engine) that
    enforces a global functional dependency upon a relation (not a rule). With a choice construct, algorithms such as worklists can be expressed effectively and efficiently. 
    \item 
    We show that the semantics of relation-based choice is easily implementable in an engine like Souffl\'e with its intermediate representation, called the Relational Algebra Machine (RAM).
    \item We explain the differences between the semantics of rule-based choice in prior work~\cite{Giannotti91} and relation-based choice in Souffl\'e. We demonstrate that relation-based choice is easier to understand by users of large-scale Datalog programs.
\end{itemize}

\section{Motivating Example}\label{sec:motivating-example}
Compilers and productivity tools require worklist algorithms~\cite{worklist02}, especially for 
control and data-flow analysis~\cite{dragonbook}. 
As part of more elaborate analyses, an example for a worklist algorithm is the construction 
of a spanning tree of a control-flow graph. This kind of application can be 
found for efficient placement of profiling code in programs~\cite{profiling96},
dataflow analysis~\cite{hecht1975simple,sharir1981strong}, 
and loop reductions~\cite{hecht1974characterizations}.

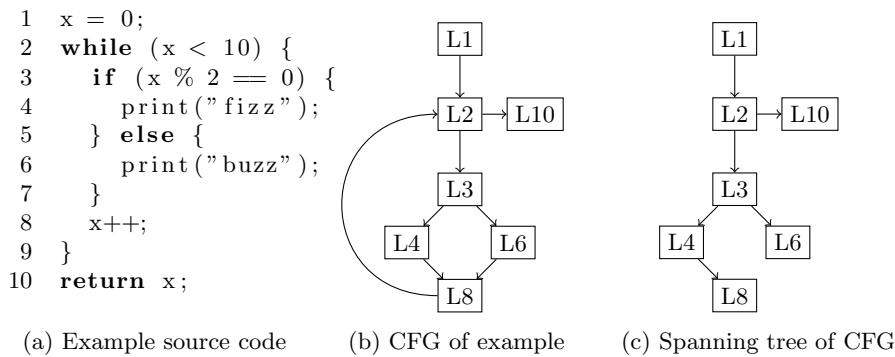
\begin{figure}[t]
\centering
\begin{subfigure}[b]{0.32\textwidth}
\begin{lstlisting}[language=c, numbers=left, basicstyle=\footnotesize,%
        tabsize=1, stepnumber=1, xleftmargin=5.0ex]
x = 0;
while (x < 10) {
  if (x % 2 == 0) {
    print("fizz");
  } else {
    print("buzz");
  }
  x++;
}
return x;
\end{lstlisting}
\caption{Example source code} \label{fig:example-code}
\end{subfigure} 
\begin{subfigure}[b]{0.32\textwidth}
    \begin{tikzpicture}[ el/.style = {inner sep=2pt, align=left, sloped},
    every label/.append style = {font=\tiny}
    ]
        \node[shape=rectangle, draw=black] (L1) {L1};
        \node[shape=rectangle, draw=black, below of=L1] (L2) {L2};
        \path [->](L1) edge (L2);
        
        \node[shape=rectangle, draw=black, below of=L2] (L3) {L3};
        \path [->](L2) edge (L3);
        \node[shape=rectangle, draw=black, below left of=L3] (L4) {L4};
        \path [->](L3) edge (L4);
        \node[shape=rectangle, draw=black, below right of=L3] (L6) {L6};
        \path [->](L3) edge (L6);
        \node[shape=rectangle, draw=black, below right of=L4] (L8) {L8};
        \path [->](L4) edge (L8);
        \path [->](L6) edge (L8);
        \path [->, bend left=90, looseness=1.8](L8.west) edge (L2.west);
        \node[shape=rectangle, draw=black, right of=L2] (L10) {L10};
        \path [->](L2) edge (L10);

    \end{tikzpicture}
    
    \caption{CFG of example} \label{fig:example-cfg}

\end{subfigure}
\begin{subfigure}[b]{0.32\textwidth}
    \hspace*{18pt}\begin{tikzpicture}[ el/.style = {inner sep=2pt, align=left, sloped},
    every label/.append style = {font=\tiny}
    ]
        \node[shape=rectangle, draw=black] (L1) at (4.0, 2.0) {L1};
        \node[shape=rectangle, draw=black, below of=L1] (L2) {L2};
        \path [->](L1) edge (L2);
        
        \node[shape=rectangle, draw=black, below of=L2] (L3) {L3};
        \path [->](L2) edge (L3);
        \node[shape=rectangle, draw=black, below left of=L3] (L4) {L4};
        \path [->](L3) edge (L4);
        \node[shape=rectangle, draw=black, below right of=L3] (L6) {L6};
        \path [->](L3) edge (L6);
        \node[shape=rectangle, draw=black, below right of=L4] (L8) {L8};
        \path [->](L4) edge (L8);
        \node[shape=rectangle, draw=black, right of=L2] (L10) {L10};
        \path [->](L2) edge (L10);
    \end{tikzpicture}
    \caption{Spanning tree of CFG} \label{fig:example-st}
\end{subfigure}
\caption{Running Example, showcasing a snippet of source code with the corresponding control flow graph and spanning tree} 
\label{fig:running-example}
\end{figure}

Control flow graphs (CFGs) express the traversal of control in a program whose nodes are basic blocks (linear code) and edges of the graph indicate potential traversal between two 
basic blocks. Fig.~\ref{fig:example-code} shows an input program whose control flow is depicted in Fig.~\ref{fig:example-cfg}. The nodes in the control-flow graphs refer to the statements in the corresponding lines of Fig.~\ref{fig:example-code}. The spanning tree of the CFG is illustrated in Fig.~\ref{fig:example-st}, containing all the nodes of the 
CFG, but with only a subset of edges. Each node has at most one incoming edge and all nodes are connected, thus forming a spanning tree.

A standard worklist algorithm to compute a spanning tree is shown 
in Fig.~\ref{fig:worklist-st}. A worklist contains all the nodes 
that ought to be visited in the next few iterations. 
The set \texttt{nodes} is used to store all 
visited nodes so far. The set \texttt{st} 
is used to store the edges of the spanning tree. The worklist is 
initialized with the root node, an artificial node with no incoming edge and a single out-going edge to the first basic block of the program.  
New nodes of the spanning tree are discovered and added to the worklist in each iteration, until 
no more valid nodes exist and the worklist becomes empty.
Inside the loop, the worklist algorithm chooses an arbitrary node 
from the worklist. For this 
node, all adjacent nodes that haven't been visited
yet will be added to the worklist and the spanning 
tree edges are constructed for the newly discovered 
nodes. With the worklist algorithm we can discover 
all reachable nodes and build the spanning tree 
in the discovery process. 

While existing Datalog systems can be effectively used for many modern program analysis workloads~\cite{jordan2016souffle,bravenboer2009strictly}, worklist-style algorithms  are often challenging. Since standard modern Datalog engines are deterministic, they must explore \emph{all} paths in a graph to compute a spanning tree, before making an arbitrary choice using a complex induction procedure. 
\textit{Datalog}~\cite{Abiteboul1995FoundationOfDB} 
represents programs as Horn clauses of the 
form $L_{0}\ \logicTurnstile\ L_{1},$ $\ldots,$ $L_{n}$.
Each $L_{i}$ has the form $R_{i}(x_{1},\ \ldots,\ x_{m})$; we say $L_i$ is a
\emph{predicate} with relation $R_i$ of arity $m$, and each attribute $x_i$ is ether a constant or a variable. 
When the right hand side (the \emph{body}) is empty, the Horn clause is
interpreted as a fact; facts are unconditionally true.
Otherwise, the Horn clause is interpreted as a rule, which means the head of the clause 
is true when
all the literals in the body are evaluated to true: $L \logicTurnstile\ L_1,\ \ldots,\ L_n. $
In particular, stratified negation~\cite{Abiteboul1995FoundationOfDB}, which is a standard semantics in Datalog to handle negation, does not permit a straightforward implementation of the worklist-style algorithms.

For example the spanning tree algorithm could be implemented with a rule such as  \verb?st(v,u)? \verb?:-? \verb?st(_,v),? \verb?edge(v,u),? \verb?!st(_,u)?.
However, this is illegal in standard 
Datalog engines because 
it contains a negation that is not stratified~\cite{Abiteboul1995FoundationOfDB}, i.e., 
the recursive relation \texttt{st} depends on the negation of \texttt{st} itself. 
The \emph{choice} construct for rules overcomes the problem of 
choosing elements~\cite{initialchoice},
which also improves the overall expressive power of Datalog programs~\cite{expressivepower}. 
In this work, we introduce a variation of rule-based choice which we call a relation-based choice.  Consider the spanning tree example expressed in the Souffl\'e language as illustrated in Fig.~\ref{fig:spanning-tree}. 
The Datalog program  imposes a functional dependency constraint for 
relation \texttt{st} with the keyword \texttt{choice-domain} on 
attribute \texttt{u}. 
The functional dependency constraint ensures that for a given value of attribute $u$ there exists at most one tuple. For example, if the relation \texttt{st} already contains the tuple $(\texttt{L5},\texttt{L9})$,  
a subsequent insertion of a tuple such as $(\texttt{L7}, \texttt{L9})$ whose \texttt{u}'s attribute value is $\texttt{L9}$
will be suppressed. 
With that functional dependency, the relation \texttt{st} becomes a function 
whose domain is the attribute domain of $v$ and its co-domain is the attribute domain of $u$. For sake of brevity, we omit the co-domain declaration in Souffl\'e so that all the excluded attributes of the domain specification implicitly become the attributes of the co-domain.

\begin{figure}[t!]
\centering
\captionsetup[subfigure]{position=b}
\begin{subfigure}[b]{0.38\textwidth}
    \begin{algorithmic}
        \STATE $worklist \gets \{root\}$ 
        \WHILE {$worklist \neq \emptyset$}
            \STATE $u \gets$ a choice from $worklist$
            \STATE $nodes \gets nodes \cup \{u\}$ 
            \FOR {$v$ in $adj(u) \setminus nodes$}
                \STATE $st \gets st \cup \{(u, v)\}$
                \STATE $worklist \gets worklist \cup \{v\}$
            \ENDFOR
        \ENDWHILE
    \end{algorithmic}
\caption{Worklist Algorithm} \label{fig:worklist-st}
\end{subfigure} 
\hfill
\begin{subfigure}[b]{0.58\textwidth}
\begin{lstlisting}[basicstyle=\footnotesize, tabsize=1, stepnumber=1, basicstyle=\ttfamily, columns=fullflexible]
.decl edge(v:symbol, u:symbol)
.input edge
.decl st(v:symbol, u:symbol) choice-domain u
.output st
st("root","L1"). 
st(v,u) :-  st(_, v),  edge(v,u).
\end{lstlisting}
\caption{Souffl\'e with Choice} \label{fig:worklist-souffle}
\end{subfigure}
\caption{Spanning Tree: Worklist Algorithm vs Souffl\'e with Relation-based Choice}
\label{fig:spanning-tree}
\end{figure}

Without a choice construct, 
the notion of non-deterministic choice must be simulated via induction. This process is quite complex due to stratified negation.
Stratification ensures that a simple expression of a complement set (i.e., to eliminate nodes that have already been visited) is impossible, since doing so would involve a non-stratified negation. Instead, an algorithm written in stratified Datalog must construct an explicit complement relation, and use induction to select the next valid edge. Thus, while a spanning tree algorithm is expressible in modern Datalog engines (see Appendix~\ref{sec:appendix-spanning-tree} for a Souffl\'e implementation), the native solution is very expensive in terms of runtime, memory usage, and code complexity. 

To describe the native implementation in more detail, a rooted spanning tree is built incrementally from a chosen start node. The program repeatedly adds individual valid edges into the graph until no edges can be added. Since several edges may be valid at any given point, and we wish to explore only one arbitrary path, we must adorn the input edges with a total order so that ties among incoming edges can be broken. As the ordering is arbitrary, it is enough to assign a unique identifier to each edge in the graph. In Souffl\'e, unique numbers can be generated using the global counter, \texttt{\$}, a unary functor which generates numbers sequentially when used, starting from the number zero (line~\ref{line:cnt}). 
After creating an order among edges,  an induction chooses the next 
valid edge from the worklist. A single valid edge must be chosen
in each step, with elements with a lower ID being prioritized 
to break ties. We introduce a helper relation \verb?chosenEdgeInductive? (line~\ref{line:chosen}) 
with attribtues \verb?step?, \verb?edge_id? and \verb?is_chosen?
for constructing the induction.
The \texttt{step} number identifies the current state of construction, incrementing with each new edge added into the spanning tree. For each step, we seed the induction with a dummy base case. The recursive rule then sequentially checks every edge, incrementing the edge ID being checked while they remain invalid. As soon as a valid edge is found, it is selected, and the recursive case terminates. A tuple in the relation contains a \texttt{TRUE} in the final column if and only if the edge with the given edge ID was chosen at that step.
We cannot simply negate \texttt{validEdge} to check if an edge is invalid in the recursive rule for \texttt{chosenEdgeInductive}, since the validity of an edge relies on the choices made in previous steps, which in turn depends on this inductive rule again. Therefore, the assumptions of stratified negation would be broken. Instead, \texttt{invalidEdge} must be constructed positively alongside \texttt{validEdge}.

The resulting program requires deeply recursive rules using inductive arguments, the notion of total orders, and the positive construction of  complement sets.
Hence, the simulation of choice in logic 
is tedious and error-prone resulting in programs 
with sub-optimal performance.
In contrast, the choice construct enables a much simpler and far more efficient expression of a spanning tree algorithm. In contrast to the 21 Datalog rules required for the native Souffl\'e implementation, the running example in Fig.~\ref{fig:example-st} demonstrates an implementation with 1 rule and a choice constraint for the relation \texttt{st}.

\section{Semantics of Choice} \label{sec:choice-theory}

In the previous section, we established that a 
choice construct in a language like Souffl\'e 
is fundamental for implementing worklist 
style algorithms. However, there are two 
options for implementing choice in a Datalog
engine. The choice construct can be either (1) rule-based
or (2) relation-based.  In this section, we first explain the semantics of 
relation-based choice, which we choose to implement in Souffl\'e. 
We then explain the slight differences between the semantics of 
relation-based choice and rule-based choice. 
After that, we provide an example demonstrating why we believe relation-based choice
makes more sense in modern Datalog language.
Finally, we show that the expressive power of two different choice constructs 
are really the same and 
how to simulate rule-based choice semantics with relation-based choice construct.

\paragraph{Relation-based Choice.} Relation-based choice extends the expressiveness of logic languages (e.g., Datalog) by introducing non-determinism into the logic framework at the relation level. In particular, choice constraints are declared for a relation, allowing programs to arbitrarily make a single choice out of a set of possible candidates. For example, a relation declared with choice constraints in the Souffle Language has the form:
\begin{align*}
    \texttt{.decl}\ A(X_1,\ldots,X_n)\ \textcolor{blue}{\texttt{choice-domain} \ D_1, \ldots, D_k}
\end{align*}
Here, $A$ is the relation name, and the 
sequence $X_1, \ldots, X_n$ forms the 
attributes of the relation. The choice 
constraints, $\texttt{choice-domain} \ D_1, \ldots, D_k$ imposes 
a set of relation-level constraints on the relation, where each 
domain ${D_i}$ is a subset of attributes of the relation
$D_{1}, \ldots, D_{m} \subseteq \{X_1, \ldots, X_n\}$. 
For example, a relation \texttt{A} declared with 
\texttt{.decl A(x:number, y:number, z:number) choice-domain x, (x,z).} 
has to respect two functional dependencies: 
$x \rightarrow (y, z)$ and $(x, z) \rightarrow y$.
Semantically, each choice constraint $D_i$ encodes a 
relation-level invariant which ensures that there is 
at most a single tuple in the relation for any particular 
value for the attributes in the choice domain. This 
constraint is similar to the notion of primary or 
candidate keys in a relational database~\cite{DBConstraints}.

We extend the standard fixpoint semantics of Datalog \cite{Abiteboul1995FoundationOfDB}. The choice construct must have the ability to arbitrarily \emph{choose} tuples in a relation such that the resulting set of tuples satisfies the choice constraint. Consider a relation $A$ with attributes $X_1, \ldots, X_n$. Let $D \subseteq \{X_1, \ldots, X_n\}$ be a choice domain, let $M_A$ be the Cartesian product of the attribute domains of $A$, let $\mathcal{A} \subseteq M_A$ be a set of tuples for $A$, and let $\mathcal{A} \big|_{D}$ be the set of instantiated values when tuples in $\mathcal{A}$ are restricted to $D$. 
A \emph{choice function} $c_D : 2^{M_A} \rightarrow 2^{M_A}$ on a set of tuples, $\mathcal{A}$, for the relation $A$ can be defined as 
\begin{align*}
    c_D (\mathcal{A}) := \left \{\textsf{SingleChoice} \left (\{t \in \mathcal{A} \mid t \big|_{D} \in \mathcal{A} \big|_{D}\} \right) \right\}
\end{align*}
where $t \big|_{D}$ is the set of instantiated values for attributes in $D$ for the tuple $t$. For each instantiation of attributes $X_i$ in $D$, $c_D$ chooses exactly one tuple matching that instantiation (via an extra function $\textsf{SingleChoice}$ that arbitrarily chooses one element in the set). In other words, the choice function enforces uniqueness of values in the choice domain by arbitrarily choosing one tuple matching each instantiation. If $M$ is the Cartesian product of the domain of relations in Datalog program $P$, then the choice function can be extended as $c : 2^M \rightarrow 2^M$, which applies $c_D$ to each relation with choice constraints.
The result of applying the choice function $c$ to a Datalog instance is an instance that satisfies the uniqueness condition of the choice constraints, by arbitrarily choosing one tuple for each instantiated set of values for each choice domain.

The other important semantics for choice constraints is to exclude tuples that already define values for the choice domain. The exclusion semantics applies for recursive rules, where an earlier iteration may define some values for the choice domain, while a later iteration computes the same values. In this situation, the tuples in the later iteration should be rejected, since those values in the choice domain are already chosen. Given another set of tuples $\mathcal{A'}$, the instantiations in $D$ that are already defined in $\mathcal{A}$ can be excluded by the exclusion function:
\begin{align*}
    e^\mathcal{A}_D (\mathcal{A'}) := \mathcal{A'} \setminus \{t \in \mathcal{A'} \mid t \big|_{D} \in \mathcal{A} \big|_{D}\}
\end{align*}
The exclusion function can also be extended to an instance $I$, where $e^I (I')$ applies exclusion for the whole instance, excluding tuples in $I'$ where values for the choice domain are already defined in tuples in $I$.

We extend the standard semantics of Datalog with choice constraints such that the result of applying the consequence operator always satisfies these constraints (using bottom-up evaluation). For this, we define a \emph{choice consequence operator}, $\Gamma^c_P$, which applies the exclusion and choice operations, to $I$ as follows:
\begin{align*}
    \Gamma^c_P (I) = I \cup c(e^I(\{t \mid t \logicTurnstile\ t_1, \ldots, t_k \text{ is a rule instantiation with each } t_i \in I\}))
\end{align*}
It can be seen that $\Gamma^c_P (I)$ is monotone. Therefore, we can show that there exists a minimum fixpoint of $\Gamma^c_P (I)$ by using \emph{Tarski's Fixpoint Theorem} \cite{tarski1955}. The resulting fixpoint is denoted the \emph{choice constraint model} of Datalog program $P$ given instance $I$.

To evaluate a Datalog program containing relation-based choice constraints, the standard semi-na\"ive evalution algorithm (i.e., Algorithm~\ref{alg:semi-naive} introduced in Appendix \ref{sec:appendix-seminaive-evaluation}) can be extended using a semi-na\"ive version of the choice consequence operator. We can define such an operator in a similar fashion to the standard choice consequence operator, by applying the choice and exclusion functions:
\begin{align*}
    \Gamma^c_P (\Delta, I) = I \cup c\left( e^I \left(\left\{t\ \middle\vert \begin{array}{l} t \logicTurnstile\ t_1, \ldots, t_k \text{ with each } t_i \in I \\ \text{and at least one } t_j \in  \Delta \end{array} \right\}  \right) \right)
\end{align*}
Using a semi-na\"ive choice consequence operator, Algorithm~\ref{alg:semi-naive} in Appendix \ref{sec:appendix-seminaive-evaluation} should be modified such that line~\ref{line:semi-naive-rule-eval} uses $\Gamma^c_P$ instead of $\Gamma_P$. With this simple change, the efficient fixpoint evaluation of a choice program can be achieved.

\paragraph{Rule-based Choice.} Unlike relation-based choice, rule-based choice from prior work enforces 
the functional dependency on the rule level. 
That is, only the tuples generated by the rules 
with the choice constructs have to respect the functional dependencies.
Let's consider the rule-based choice 
version of the rooted spanning tree as an example. 
\begin{verbatim}
st("root","L1"). 
st(v, u) :- st(_, v), edge(v,u), choice((u), (v)).
\end{verbatim}

The keyword \texttt{choice((X), (Y))} 
specifies the functional dependency $X \rightarrow Y$ on the rule-level.
Unlike the relation-based implementation, only the second rule in the above
program has to respect the functional dependency,
while the resulting relation \texttt{st} can still have 
a non-injecting relation between  $X$ and $Y$.
In fact, the above program does not work as intended. 
Although the choice construct on second rule enforces that every end node $u$ has
a unique predecessor, there is nothing preventing the second rule from generating
another edge to the starting node \texttt{L1}.
This does not break the functional dependency 
because constraint is only enforced on rule-level and 
the tuple \texttt{st("root", "L1")} was specified in another clause in line one.
To correct this, we need to rewrite the second rule as
\begin{verbatim}
st(v, u) :- st(_, v), edge(v,u), choice((u), (v)), u != "L1".
\end{verbatim}
This program demonstrates a classic example where rule-based choice semantics 
can sometime become error-prone 
and hard to handle in large scale Datalog programs where each relation has dozens of rules.

\paragraph{Expressive Power.} Although the user experience may differ, rule-based choice and relation-based choice have the same expressive power.
We present an example of rewriting the rooted spanning tree example using rule-based choice semantics, but using relation-based choice construct.
Consider the semantics of the rule-based choice implementation given under the stable model:
\begin{verbatim}
st("root","L1"). 
st(v, u) :- st(_, v), edge(v,u), chosen(u, v), u != "L1". 
chosen(u, v) :- st(_, v), edge(v, u), !diffChoice(u, v). 
diffChoice(u, v) :- chosen(u, v’), v != v'.    
\end{verbatim}
The above program cannot be computed under stratified semi-naive evaluation
because of the cyclic negation between \texttt{chosen} and \texttt{diffChoice}.
However, it is given by Giannotti et al.\cite{Giannotti91,Giannotti04} under
the stable model to formally describe the semantics of the rule-based choice implementation.
The intuitive meaning of the program is to use an auxiliary table (\texttt{diffChoice}) to record
the generated tuples and prevent the rule from generating tuples that violate the dependency.
The implementation given by Giannotti et al. follows this intuition, and uses an auxiliary table internally.
To mimic the effect of this with relation-based choice, we use a separate relation \texttt{st'} with a relation-based choice constraint to act as the auxiliary table.
\begin{verbatim}
.decl st'(v:symbol, u:symbol) choice-domain(u)
.decl st(v:symbol, u:symbol) 
st("root","L1").
st'(v, u) :- st(_, v), edge(v,u), u != "L1".
st(v, u) :- st'(v, u).
\end{verbatim}
In Section~\ref{sec:souffle-implementation} we show that because of how relation-based choice is implemented, this emulation does not
suffer from any extra overhead and has the exact same cost as the one proposed in
the literature where an auxiliary table is used.

\section{Implementation in Soufflé}\label{sec:souffle-implementation}

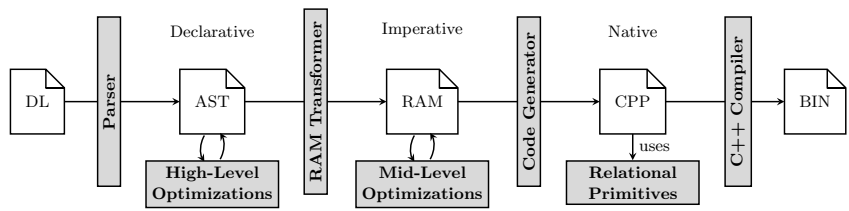
\begin{figure}[H]
\centering
\hspace*{-1cm}
\scalebox{0.7}{
\input{souffle-pipeline}
}
\caption{Execution model of Souffl\'e.}
\label{fig:Souffle-model}
\end{figure}

In the following, we describe the implementation of relation-based choice in the state-of-the-art Datalog engine Souffl\'e~\cite{jordan2016souffle}.
A general overview of the Souffl\'e infrastructure is shown in Fig.~\ref{fig:Souffle-model}.
Souffl\'e parses the input Datalog program into an Abstract Syntax Tree (AST) representation. 
After parsing, Souffle applies a series of high-level optimizations on the AST representation. The AST contains information including all declared relations, rules and facts of the source program. After applying the AST optimisations, the AST representation is lowered into an intermediate representation called the Relational Algebra Machine (RAM). A RAM program consists of a set of relational operations along with imperative constructs. Mid-level optimizations are then applied to the RAM code, which finally is synthesized into an equivalent C++ program (or is interpreted). 

A relation can be declared with zero or more choice constraints, each of which can contain a single attribute or a list of attributes. 
We extend the Souffl\'e parser to read a list of choice domains, written in the same form as shown in Section~\ref{sec:choice-theory}. We extend the current representation of relations in Souffle's with an extra attribute, storing each choice-domain as a list of indices representing the corresponding attributes' positions in the relation. For example, a relation declaration \texttt{.decl A(x, y) choice-domain x} will have a single choice-domain value $\{0\}$ denoting that the first attribute in $A$ is in the choice-domain. 
A semantic check ensures that each choice-domain is valid (i.e., the attributes appear in the source relation), and a high-level optimization is used to reduce any redundant constraints.

At the final stage and before execution, we insert extra RAM operations to ensure the semantic for each insertion happens on a relation with choice-domain.
We have various RAM elements  implementing the semantics:

\begin{enumerate}
    \item \texttt{TupleElement(t,i)} (or simply \texttt{t[i]}): It takes a runtime tuple $t=(t_1, \ldots, t_n)$ and an index $i$ as arguments, and returns the value of the $i^{\text{th}}$ element of $t$.
    \item \texttt{Insert(t,R)}: It inserts a runtime tuple $t = (t_1, \ldots, t_n)$ into a relation $R$.
    \item \texttt{ExistenceCheck(P,R)}: It checks if the given pattern $P=(p_0, \ldots, p_n)$ exists in the relation $R$. $p_i$ can be either a runtime expression (e.g., \texttt{TupleElement}), a constant or a special value $\bot$ which matches with any value.
\end{enumerate}

While inserting a tuple $t$ into a relation $R$, 
the RAM program checks whether a choice constraints is violated. For this, we apply the choice function $c$ and the exclusion function $e^I$ mentioned in section \ref{sec:choice-theory}.
Before an \texttt{Insert(t,R)} operation (which would add tuple \texttt{t} to relation \texttt{R}), we add an extra
check \texttt{ExistenceCheck(P,R)} with pattern $P=(p_0, \ldots, p_n)$. The value of $p_i$ is defined as:
\begin{equation*}
p_i = 
\begin{cases}
\texttt{TupleElement(t, i)} & \text{if the $i^{th}$ attribute is in $D$ } \\
\bot  & \text{otherwise}\\
\end{cases}       
\end{equation*}
where $D$ is the  choice-domain $D=\{d_0, \ldots, d_k\}$ on $R$.
If the existence check finds a matching tuple, then the insert operation is rejected. Thus, prior to the insertion tuples are filtered so that only tuples 
that do not violate the functional dependency constraint of the choice domain are inserted.

For a non-recursive rule, the relation $R$ in the \texttt{ExistenceCheck} would be the original relation that the tuple is inserted into. However, for a recursive rule, the relation $R$ would denote a \emph{new} auxiliary relation rather than the original one (for semi-nai\"ve evaluation), which requires the exclusion function. To achieve this in RAM, a similar existence check is applied to each version of the relation, i.e., if $R$ has the form \texttt{R'}, then we also create an \texttt{ExistenceCheck(P,R')}, which ensures that any new tuples inserted into the relation will not replicate values for the choice-domains already defined in an earlier iteration, thus executing the semantics of the exclusion function.

\begin{figure}[t]
\small
\begin{algorithmic}
\STATE \textbf{Input:} AST representation of the source program.
\STATE \textbf{Output:} RAM representation with insertion guarded by existence check to guarantee the choice domain.
\STATE $RAM \gets$ translate the AST into RAM without concerning choice
\FOR{each insertion \texttt{Insert(t, R)} in $RAM$}
    \IF{$R$ has choice-domain}
        \STATE $G \gets$ a new \texttt{GuardedInsert(t, R, E=$\emptyset$)}
        \FOR{each choice-domain $C$}
            \STATE add \texttt{ExistenceCheck(P, R)} into $E$, $p_i = t[i]$ if $i \in C$ else $p_i = \bot$
        \IF{$R$ has prefix \texttt{NEW}}
            \STATE $R' \gets$ the corresponding original relation of $R$.
            \STATE add \texttt{ExistenceCheck(P, R')} into $E$.
        \ENDIF
        \ENDFOR
    \STATE replace the existing insertion with $G$ in $RAM$.
    \ENDIF
\ENDFOR
\STATE return $RAM$
\end{algorithmic}
\caption{Augmenting a RAM program with Guarded Insertions} \label{alg:choice-translator}
\vspace{-12pt}
\label{guardedProjectg}
\end{figure}

\begin{figure}[ht]
\centering\small
\begin{lstlisting}[mathescape=true, basicstyle=\footnotesize, tabsize=1, %
        stepnumber=1, basicstyle=\ttfamily, columns=fullflexible, escapeinside={<@}{@>}]
INSERT ("root", "a") INTO new_st
READ INPUT INTO delta_st.
LOOP
  IF ((NOT (delta_st = $\emptyset$)) AND (NOT (graph = $\emptyset$)))
    FOR a IN delta_st
      FOR b IN graph ON INDEX b[0] = a[1]
        <@\textcolor{blue}{IF (NOT ($\bot$,b[1]) $\in$ new\_st) AND (NOT ($\bot$,b[1]) $\in$ st)}@>
          INSERT (a[1], b[1]) INTO new_st
   BREAK IF (new_st = $\emptyset$)
   MERGE new_st INTO st
   SWAP (delta_st, new_st)
   CLEAR new_st
END LOOP
\end{lstlisting}
\vspace{-14pt}
\caption{Resulting RAM program from spanning tree with relation-based choice}
\vspace{-16pt}
\label{fig:st-ram}
\end{figure}

To encapsulate the semantics of the filtering insertions, we introduce a new RAM operation, \texttt{GuardedInsert(t, R, E)}, i.e., a regular \texttt{Insert} operation with an extra field $E$ representing a list of \texttt{ExistenceCheck} operations. 
The semantics of \texttt{GuardedInsert} specifies that the insertion only proceeds if all existence checks in $E$ have been done. 
An algorithm is given in Fig.~\ref{alg:choice-translator}, demonstrating the process of translating a Souffl\'e program with choice constraints. 
In this algorithm each existing \texttt{Insert} operation is translated into a corresponding \texttt{GuardedInsert} operation, which encodes the semantics of the choice and exclusion functions.

With the new RAM transformation, the spanning tree program (Fig.~\ref{fig:worklist-souffle}) is translated into the RAM program as shown in Fig.~\ref{fig:st-ram}. The parts highlighted in blue are the extra existence checks introduced by the new translator. 
Because relation-based choice only requires extra existence checks , it is easy to see the emulation we describe in section~\ref{sec:choice-theory} has the same cost as the rule-based choice implementation proposed in prior work.

Souffl\'e is equipped with highly-efficient data structures such as the specialized B-tree~\cite{jordan2019btree}. 
During the translation from RAM to C++, Souffl\'e analyzes the RAM representation to automatically compute indices for each primitive search~\cite{subotic2018indexselection}. This automatic index selection allows Souffl\'e to generate static C++ code that is tailored to data structures specialized for each index. As a result, the existence checks can be done efficiently with minimal overhead.

\section{Experiments}
This section explores the performance benefit of choice construct in
Souffl\'e compared to native Souffl\'e without choice, as well as exploring any performance difference between relation-based choice and rule-based choice.
Our experimental results illustrate that both choice constructs improve the environment of native Souffl\'e with similar performance statistics. 
Furthermore, we also demonstrate the applicability of choice and how it extends the expressive power of logic language. These experiments aim to answer three main research questions:
\begin{enumerate}[label=\textbf{\arabic*}.]
    \item Does choice substantially improve runtime and memory performance over equivalent non-choice Datalog programs?
    \item Does choice allow for easier expressivity for Datalog programs requiring non-determinism?
    \item Is there any performance difference between relation-based and rule-based choice?
\end{enumerate}

Our experiments demonstrate a rooted spanning tree implementation applied on real-world input, along with 5 other algorithms that utilize choice constructs. For each algorithm, three versions are implemented:
\begin{enumerate}[label=\textbf{\arabic*}.]
    \item \textbf{Relation-based Choice:} a Souffl\'e program that uses relation-based choice constraint (as implemented in Section \ref{sec:souffle-implementation})
    \item \textbf{Rule-based Choice:} a Souffl\'e program that uses relation-based choice construct to emulate the rule-based choice semantics as described in Section~\ref{sec:choice-theory}.
    \item \textbf{Native:} a Souffl\'e program that uses aggregates and auxiliary relations to emulate the effects of choice without using an explicit choice constraint
\end{enumerate}

The experiments were conducted on a machine with an AMD Ryzen 2990WX 32-Core CPU and
126 GB of memory. All programs were run in sequential mode.
Both runtime and memory usage were measured using the GNU \texttt{time}
utility, observing both user time and maximum resident set size respectively.

\subsection{Rooted spanning tree} \label{sec:experiment-spanning-forest}
We extract Control Flow Graphs (CFGs) from the real-world benchmark suite \texttt{SpecCPU2000}~\cite{SPECCPU}. These CFGs consist of large graphs with small connected components, thus the spanning forest consists of one spanning tree for each connected component.
Computing the spanning tree of a program's CFG is very important
for program analysis tools to identify loops, possible
optimization opportunities and security flaws, etc.
Since each input file contains several connected
components, we modify the rooted spanning tree example in
Fig.~\ref{fig:worklist-souffle} by computing a spanning forest with relation-based choice construct:

\begin{lstlisting}[basicstyle=\footnotesize, tabsize=1, stepnumber=1, basicstyle=\ttfamily, columns=fullflexible]
.decl edge(module:symbol, x:symbol, y:symbol)
.input edge
.decl startNode(module:symbol, x:symbol)
.input startNode
.decl st(module:symbol, x:symbol, y:symbol) choice-domain (module, y)
.output st

st(M,X,Y) :- startNode(M,X), edge(M,X,Y).
st(M,X,Y) :- st(M,_,X), edge(M,X,Y).
\end{lstlisting}
The attribute \texttt{module} identifies the name of the function where each
connected component is generated from.
By providing a single root node \texttt{startNode} for each component (line 4), we compute the spanning forest for the whole graph.
The choice domain of relation \texttt{st} is specified as \texttt{(module, y)}, so that each module (connected component) contains a single spanning tree.
Finally, the rule on line 9 states that a spanning tree edge from $X$ to $Y$ in the connected component $M$ exists if the
spanning tree reaches node $X$ and there is an edge from $X$ to $Y$.

The translated rooted spanning tree program in native Souffl\'e uses an inductive approach as in Section~\ref{sec:motivating-example} and is modified in a similar way to calculate the spanning forest. 
Its implementation follows concepts from typical worklist algorithms, incrementally generating the set of edges corresponding to a spanning tree of the input graph. The inductive process ensures that each edge appears only once in the output, and 
the output edges correspond to a tree, which contains no cycles.

During this experiment, we find no measurable runtime or memory difference between the relation-based and rule-based choice implementations. Both of them are able to finish all the benchmarks within 0.1 seconds and consume a similar amount of memory.
Compared with relation-based choice, rule-based choice implementation requires an extra relation to keep track of the inserted tuples, and an extra insertion to dump the result from the auxiliary relation into the final result. However, in real-world use cases, because of the functional dependency constraint, the auxiliary relation tends to have a relatively small size, which makes the extra overhead small in comparison to the overall runtime and memory consumption.
Specifically, in this experiment, the auxiliary relation in the rule-based choice version contains only the edges of the result spanning tree, which is much smaller than the overall graph size. 
Thus, we calculated a speed-up factor based on two choice implementations to demonstrate the performance difference between the choice constructs and native Souffl\'e implementation in Table~\ref{Spec2000}.

\begin{table}[t]
    \centering\small
    \begin{tabular}{lcccccc}
    \toprule
     & \multicolumn{2}{c}{Benchmark Information} & \multicolumn{2}{c}{Runtime (seconds)} & \multicolumn{2}{c}{Memory usage (MBs)} \\
    \cmidrule(r){2-3}
    \cmidrule(r){4-5}
    \cmidrule(r){6-7}
        Program            & \shortstack{\# of \\ components} &\shortstack{average size \\ (edges)} & Native & Speedup factor & Choice & Native \\
    \midrule
        gzip     &84     &28    & 2.75 & 275.00 & 5.00 & 10.95 \\
        swim     &6      &26    & 0.02 & 2.00 & 4.72 & 5.20 \\
        applu    &16     &56    & 1.42 & 142.00 & 4.84 & 8.69 \\
        gcc      &1896   &50    & \texttt{timeout} & $>$10k & 8.00 & 573.72 \\
        art      &26     &35    & 1.57 & 157.00 & 4.93 & 9.23 \\
        equake   &26     &16    & 0.22 & 22.00 & 4.87 & 6.11 \\
        ammp     &175    &32    & 26.19 & 2619.00 & 5.14 & 28.01 \\
        sixtrack &213    &49    & 312.8 & $>$10k & 5.30 & 94.32 \\
        gap      &830    &38    & 298.2 & $>$10k & 5.84 & 116.64 \\
        bzip2    &72     &34    & 7.8 & 780.00 & 5.07 & 16.64 \\
        apsi     &96     &30    & 6.41 & 641.00 & 4.84 & 13.70 \\
        wupwise  &20     &32    & 1.7 & 170.00 & 5.02 & 9.94 \\
        mgrid    &10     &26    & 0.06 & 6.00 & 4.79 & 5.45 \\
        vpr      &261    &22    & 18.4 & 1840.00 & 5.18 & 21.84 \\
        mesa     &1064   &29    & 1258.55 & $>$10k & 5.98 & 237.61 \\
        mcf      &26     &25    & 0.26 & 26.00 & 4.82 & 6.36 \\
        crafty   &108    &88    & 1037.3 & $>$10k & 5.10 & 176.52 \\
        parser   &293    &25    & 54.79 & 5479.00 & 4.93 & 34.68 \\
        perlbmk  &234    &44    & 174.4 & $>$10k & 5.09 & 61.21 \\
        vortex   &918    &29    & 426.92 & $>$10k & 5.66 & 112.27 \\
        twolf    &180    &62    & 419.25 & $>$10k & 5.12 & 96.55 \\
    \bottomrule                     
    \end{tabular}
    \vspace{1em}
    \caption{Performance result from Spec CPU2000, timeout set to be 30 minutes.}
    \vspace{-18pt}
    \label{Spec2000}
\end{table}

The results show a significant improvement for the choice-based program compared to the native Souffl\'e program, performing at least 2$\times$ faster and up to more than 10k$\times$ faster on larger benchmarks such as \texttt{gcc} and \texttt{mesa}.
In terms of memory consumption, the choice version consumes considerably less memory than the native Souffl\'e version, and achieves a consistent memory usage across all benchmarks.  In comparison, the native Souffl\'e version uses significantly more memory as input size increases.
This is because the choice constraint only computes and stores edges that are included in
the spanning tree, which are generally fairly small compared to the constant overheads of executing a Souffl\'e program. On the other hand, the native version needs to store many intermediate computations and relies on a complex recursive scheme to obtain the same results.

Another consideration is the code complexity of both the choice constructs and native Souffl\'e implementation.
For this spanning tree problem, the native Souffl\'e implementation requires 21 rules with complex recursive structure.
On the other hand, relation-based choice version requires a minimum amount of code, with only 2 rules and a choice construct on the \texttt{st} relation.
Finally, for rule-based choice, two extra auxiliary rules and one extra constraint are used as described in Section~\ref{sec:choice-theory}.

\subsection{Other Applications}

Along with the spanning tree example, we present five other algorithms, most of them are classic examples of non-deterministic algorithms in Datalog~\cite{Giannotti04}:
\begin{itemize}
    \item \textbf{Eligible advisors}: 
    Choosing an advisor for each student.
    \item \textbf{Total order}: 
    Assigning an arbitrary total order over an unordered list.
    \item \textbf{Bipartite matching}: 
    Computing a matching over a bipartite graph.
    \item \textbf{More dogs than cats}: 
    Taking two sets of elements and deciding if one set contains more elements than the other one.
    \item \textbf{Highest mark in grade}: 
    Finding the highest mark in a subset of marks subject to a condition, e.g., the highest mark among students in each grade.
\end{itemize}

\begin{table}[t]
    \centering
    \begin{tabular}{|c|l|lll|lll|lll|}
    \hline
     & & \multicolumn{3}{c}{\shortstack{Relation-based \\ Choice}} & \multicolumn{3}{|c}{\shortstack{Rule-based \\ Choice}} & \multicolumn{3}{|c|}{\shortstack{Native}} \\
 \hline
   Program            & Input  & \shortstack{R\#} & \shortstack{T(s)} & \shortstack{M(MB)} & 
    \shortstack{R\#} & \shortstack{T(s)} & \shortstack{M(MB)} & \shortstack{R\#}  & \shortstack{T(s)} & \shortstack{M(MB)} \\
    \hline
    Eligible advisors  & \num{3000} & 1 & 0.01 & 5.5 & 2 & 0.01 & 5.7 & 4 & 0.11 & 13.7 \\
    Total order  & \num{2000} & 2 & 0.23 & 5.2 & 3 & 0.23 & 5.2 & 3 & 75.88 & 43.9 \\
    Bipartite matching  & \num{3000} & 1 & 2.73 & 93.2 & 2 & 2.73 & 93.2 & 15 & \texttt{timeout} & 771 \\
    More dogs than cats & \num{18000} & 3 & 4.42 & 7 & 4 & 4.42 & 7 & 1 & 0.01 & 6.7 \\
    Highest mark in grade & \num{10000} & 1 & 0.02 & 6 & 2 & 0.02 & 6.3 & 4 &0.02 & 6.3 \\
    \hline
    \end{tabular}
    \vspace{1em}
    \caption{Summary of experiment results.}
    \vspace{-19pt}
    \label{experiment-result}
\end{table}

Table~\ref{experiment-result} shows the results for the choice versions compared to the native Souff\'e implementations. 
No runtime or memory difference is discovered between relation-based and rule-based choice. The reason is exactly the same as for the rooted spanning tree experiment, the overhead of rule-based choice implementation is extremely small because of the functional dependency constraint force upon on the extra auxiliary relation.
Thus, in the followings, we discuss only relation-based choice and native implementations, unless otherwise specified.

For the majority of these benchmarks, choice constraints lead to significantly better performance than the native Souffl\'e version. This improvement can be attributed to native Souffl\'e versions usually requiring the full computation of a relation, followed by selecting a unique subset satisfying the equivalent functional dependencies as a post-processing step. On the other hand, choice constraints allow for the functional dependencies to be checked on-the-fly, thus not needing the full unconstrained relation, benefiting both memory and runtime.

The \emph{eligible advisors} example most clearly demonstrates the improvement in performance with the choice construct. Here, the relation-based choice can simply compute the student/advisor relationship with a single rule with a choice constraint on the \texttt{advisor} relation. However, the native Souffl\'e implementation must compute the full unconstrained \texttt{advisor} relation, with a unique numbering scheme to enforce a total ordering. Then, as a post-processing step, the algorithm selects a subset satisfying the choice constraint by using the total ordering (for example, by choosing the minimum value for the unique number).

Similar patterns can also be observed in the \emph{total order} and \emph{bipartite matching} examples.
These benchmarks demonstrate situations where choice constraints allow for both an easier and more effective specification of the problem.

On the other hand, the benchmark \emph{highest mark} shows a negligible performance difference.
In both implementations, an aggregation is used to summarize the highest mark of each grade and is the main performance bottleneck of the whole algorithm.
The performance benefit of the choice constraint that is used to restrict the result of the aggregation becomes insignificant.
However, the difference in number of rules (4 v.s. 1) still demonstrate the expressiveness of the choice constraint.

The only benchmark where the native Souffl\'e implementation outperformes the choice version is \emph{more dogs than cats}. 
In this example, the choice version consider building an injective function between the two set of elements, and then check if the domain covers all the codomain, if so, the size of the domain set is greater than or equal to the codomain set.
On the other hand, the native implementation takes a more straightforward approach, using a simple count aggregate to compute the sizes of the relations.

Importantly, for all examples, the choice version uses equal or less memory compared to the native Souffl\'e counterpart.
This improvement is a result of the auxiliary relations each
native Souffl\'e program utilizes to perform their computations. The difference is most evident in the \emph{total order} example, where the native Souffl\'e implementation suffers an
approximate 850\% increase in memory usage as a result of its auxiliary
relations. 

Going beyond performance results, every example is implemented more elegantly using choice constraints. For most of the benchmarks, the choice version contains less than half the number of rules of the native Souffl\'e version, and in three of the five benchmarks, the choice version contains only a single rule. While not a perfect measurement of elegance, the small number of rules indicates that the choice-based implementations are generally more succinct and easier to understand than the native Souffl\'e versions. As shown, native Souffl\'e implementations of programs requiring arbitrary choice, as in worklist algorithms, typically involve the construction of several intertwined recursive relations with their complements, in addition to inductive rules, aggregate functions, and imposed total orderings. Such substantial overhead often obscures the meaning of the program.  With the choice construct, such behavior is modeled with a simple constraint declaration. Moreover, the clearer semantics of the choice versions allows for a simpler extension and modification of the underlying program. For example, modifying the spanning tree example in Section \ref{sec:experiment-spanning-forest} to constrain over only the attribute \texttt{y} rather than the pair \texttt{(module, y)} would involve changing only the given choice constraint. In a native Souffl\'e implementation, changing these functional dependencies could involve substantial structural changes to the auxiliary relations to ensure correctness.

In the context of Souffl\'e, these experiments demonstrate a significant impact of choice constraints, both in terms of performance overhead as well as the ease in expressing these algorithms. Thus, the introduction of choice constraints can be seen as extending the effective expressive power of the language, since certain problems that were infeasible using aggregates and auxiliary relations can now be solved using choice constraints.

\section{Related Work}
In relational databases, the notion of functional dependencies~\cite{wiederhold1983database,beeri1977complete} is an important concept that allows a database designer to encode certain uniqueness properties as an invariant on a relation. These invariants are enforced when the relation is modified, with the database system rejecting any data that violates the uniqueness constraint.
In logic programming, a deterministic computation is expressed as a set of logic rules. To extend the capabilities of this framework, previous work has introduced the choice construct~\cite{initialchoice,NaqviT89} as a means of 
supporting non-determinism in Datalog, by enforcing uniqueness constraints similar to functional dependencies. There is some prior work on choice 
for Prolog~\cite{10.5555/1286760.1286790}.
Over the years, the applicability of choice has extended into the expression of greedy algorithms~\cite{choice,GrecoZ01,GrecoZG92}, as well as improving the overall expressive power of Datalog queries~\cite{staticexpressivepower,textbook,expressivepower}. 
It has been cited to be particularly powerful when defining aggregate 
functions for relations, especially when used in conjunction with other 
predicates~\cite{choiceaggregates}. 

Choice constructs in prior work provide an intuitive foundation for enforcing non-determinism using a rule-based choice constraint, which is applied to a singular rule in the program, so that the underlying functional dependency is exclusively enforced on the local level of the specific rule that the constraint is declared on. In order to enforce these rule-based dependencies, auxiliary relations (e.g., the $chosen$ relations in \cite{choice}) are required to provide an intermediate platform for computation for each rule with a constraint.
The semantics of rule-based choice can be tedious and error prone when applying on Souffl\'e's programs that consist of hundreds of rules and relations.

\section{Conclusion}
Extending the expressive power of logic languages is a pertinent research area, especially with these languages becoming increasingly used in real-world problems. While languages such as Datalog have found success in a number of areas,
worklist-style algorithms require notions of non-determinism which is currently challenging in modern Datalog engines. 
In this work, we report on implementing a choice construct 
in the Souffl\'e Language. We experiment with two flavors of the choice construct: rule-based choice (that has been reported in prior work) and relation-based choice, which 
we introduce in this work. 

We experiment with a number of classic algorithms using the two choice constructs
and show that using a choice construct significantly improves the performance, along with greater elegance in expressing
non-determinism in Datalog.
Our experiments indicate that there is a negligible performance difference between the two flavors of choice constructs. However, we show with an example that the semantics of rule-based choice can be tedious and error prone in Datalog programs with a large number 
of rules and relations. 

\bibliographystyle{splncs04}

\newpage
\appendix

\section{General spanning tree} \label{sec:appendix-spanning-tree}
\begin{lstlisting}[escapechar=|,numbers=left,stepnumber=1]
#define FALSE 0
#define TRUE 1

/* INPUT: graph, where (x,y) are directed edges */
.decl edge(x:symbol, y:symbol)
.input edge()

/* INPUT: start node */
.decl startNode(x:symbol)
.input startNode()

/****
 * GRAPH SETUP
 ****/
/* Read out the nodes */
.decl node(node_id:symbol)
node(x) :- edge(x,_) ; edge(_,x).

/* Assign an id to each edge */
.decl orderedEdge(id:number, x:symbol, y:symbol)
orderedEdge($,x,y) :- edge(x,y).  |\label{line:cnt}|

/* Keep track of the number of valid steps possible */
// Number of steps upper bounded by number of nodes in the graph
.decl validStep(step:number)
validStep(0).
validStep(step+1) :- validStep(step), step < count : node(_).

/****
 * VALID NODE DETECTION
 ****/
/* Determine the nodes in the tree at this step */
// A growing set where each chosen node is added per step
// Initially contains only the start node, then adds in each chosen node
.decl addedNode(step:number, node_id:symbol)
addedNode(0, start) :- startNode(start).

// Everything from the previous step remains added
addedNode(step+1, node_id) :-
    validStep(step+1),
    addedNode(step, node_id).

// The newly chosen node is also added
addedNode(step+1, node_id) :-
    validStep(step+1),
    chosenNode(step, node_id).

/* Determine the invalid nodes in this step */
// A shrinking set where only non-added nodes remain
// Initially contains all non-start nodes, 
// then removes any nodes that are now reachable
.decl unaddedNode(step:number, node_id:symbol)

// Only the start node is added initially
unaddedNode(0, node_id) :- node(node_id), !startNode(node_id).

// Each previously unadded, except the newly added node
unaddedNode(prev_step+1, node_id) :-
    unaddedNode(prev_step, node_id),
    chosenNode(prev_step, new_node_id),
    node_id != new_node_id.

/****
 * VALID EDGE DETECTION
 ****/
/* Determine the edges we can choose from */
// Any reachable edge that has not been used and does not form a cycle
.decl validEdge(step:number, edge_id: number)
validEdge(step, edge_id) :-
    orderedEdge(edge_id, _, node_id),
    unusedEdge(step, edge_id),
    reachableEdge(step, edge_id),
    unaddedNode(step, node_id).

/* Determine the edges reachable from our added node set */
// Any edge coming out of an added node
.decl reachableEdge(step:number, edge_id:number)
reachableEdge(step, edge_id) :-
    addedNode(step, node_id),
    orderedEdge(edge_id, node_id, _).

/* Determine the edges that have not been used yet */
// A shrinking set where only unused edges remain
.decl unusedEdge(step:number, edge_id:number)

// No edge is used initially
unusedEdge(0, edge_id) :- orderedEdge(edge_id,_,_).

// In each step, discard the newly used edge
unusedEdge(prev_step+1, edge_id) :-
    unusedEdge(prev_step, edge_id),
    chosenEdge(prev_step, new_edge_id),
    new_edge_id != edge_id.

/* Determine the edges that we cannot choose from */
// Constructed complement to validEdge
// Contains:
//  (1) edges already used
//  (2) edges coming out of invalid nodes
//  (3) edges going into added nodes
.decl notValidEdge(step:number, edge_id:number)
// (1) edges that have already been used
notValidEdge(step, edge_id) :-
    validStep(step),
    chosenEdge(prev_step, edge_id),
    prev_step < step.
// (2) edges coming out of unadded nodes
notValidEdge(step, edge_id) :-
    unaddedNode(step, node_id),
    orderedEdge(edge_id, node_id, _).
// (3) edges going to already used nodes
notValidEdge(step, edge_id) :-
    addedNode(step, node_id),
    orderedEdge(edge_id, _, node_id).

/****
 * EDGE CHOICE 
 ****/
/* Determine the next edge to be chosen in the sequence */
.decl chosenEdge(step:number, edge_id:number)
chosenEdge(step, edge_id) :-
    chosenEdgeInductive(step, edge_id, TRUE).

/* Inductive helper relation to find the next edge to choose */
// Go through our list of potential edges to choose in order until 
// we hit the one we want
.decl chosenEdgeInductive(step:number, edge_id:number, is_chosen:number) |\label{line:chosen}|

// Dummy base case for each step
// Edge ID assignment starts at 0, so dummy ID should be -1
chosenEdgeInductive(step, -1, FALSE) :- validStep(step).

// Inductive case - choose the first valid edge
chosenEdgeInductive(step, cur_edge_id, is_chosen) :-
    // Only keep going if the last edge wasn't valid
    chosenEdgeInductive(step, prev_edge_id, FALSE),
    cur_edge_id = prev_edge_id + 1,

    // Decide if it can be chosen
    ((validEdge(step, cur_edge_id), is_chosen = TRUE) ;
     (notValidEdge(step, cur_edge_id), is_chosen = FALSE)).

/****
 * NODE CHOICE
 ****/
/* Determine the next node to be added */
// Entirely based on edge choice
.decl chosenNode(step:number, node_id:symbol)
chosenNode(step, node_id) :-
    chosenEdge(step, edge_id),
    orderedEdge(edge_id, _, node_id).

/***
 * SPANNING TREE
 ****/
.decl st(x:symbol, y:symbol)
.output st
st(x, y) :- chosenEdge(_, edge_id), orderedEdge(edge_id, x, y).
\end{lstlisting}

\section{Semi-na\"ive Evaluation on Choice Constraints}
\label{sec:appendix-seminaive-evaluation}
While na\"ive evaluation provides a straightforward evaluation algorithm for Datalog, it also suffers from computational inefficiencies where previously computed results are needlessly re-computed.
To overcome these inefficiencies, semi-na\"ive evaluation was introduced~\cite{Abiteboul1995FoundationOfDB}. The intuitive idea of semi-na\"ive evaluation is that 
only new facts generated in the previous iteration need to be
considered in the current iteration.

Semi-na\"ive evaluation introduces two auxiliary relations for each recursive relation in the original program, $K_{new}$ and $\Delta$. $K_{new}$ includes the tuples which are newly generated in the current iteration, while $\Delta$ includes the tuples which were new in the previous iteration.
We then denote $\Gamma_P(\Delta, I)$ as the set of tuples that result from
applying the immediate consequence operator on an instance $I$, with newly generated facts
$\Delta$. Formally,
\begin{align*}
    \Gamma_P (\Delta, I) = I \cup \left\{t\ \middle\vert \begin{array}{l} t \logicTurnstile\ t_1, \ldots, t_k \text{ is a rule instantiation with each } t_i \in I \\ \text{and at least one } t_j \in  \Delta \end{array} \right\}
\end{align*}

Using this semi-na\"ive form of the immediate consequence operator, the general algorithm for
semi-na\"ive evaluation is shown in Algorithm~\ref{alg:semi-naive}. In this algorithm, the $\Delta$ relation is initialized to be the input EDB, while the full relations are initialized to be empty (lines~\ref{line:semi-naive-initialize-delta} and \ref{line:semi-naive-initialize-rel}). Then, in the fixpoint loop, the Datalog rules are evaluated using $\Delta$, and the results are stored in the \emph{new} version of the relation (line~\ref{line:semi-naive-rule-eval}). If $K_{new} = \emptyset$, then a fixpoint has been reached, and the algorithm exits (line~\ref{line:semi-naive-break}). Otherwise, the algorithm continues by merging $K_{new}$ into the full relation $K$ (line~\ref{line:semi-naive-merge}), then setting $\Delta$ to be $K_{new}$ in preparation for the next iteration (line~\ref{line:semi-naive-delta}).
\begin{algorithm}[h]
\caption{Semi-na\"ive evaluation} \label{alg:semi-naive}
\begin{algorithmic}[1]
\STATEx \textbf{Input:} A Datalog program $P$ and some EDB $I$.
\STATEx \textbf{Output:} A minimal fixpoint of $P$.
\STATE $\Delta \gets I$ \label{line:semi-naive-initialize-delta}
\STATE $K \gets \emptyset$ \label{line:semi-naive-initialize-rel}
\LOOP
    \STATE $K_{new} \gets \Gamma_P(\Delta, I) - K$  \label{line:semi-naive-rule-eval}
    \IF{$K_{new} = \emptyset $}
        \STATE break loop \label{line:semi-naive-break}
    \ENDIF
    \STATE $K \gets K_{new} \cup K$ \label{line:semi-naive-merge}
    \STATE $\Delta \gets K_{new}$ \label{line:semi-naive-delta}
    \STATE $K_{new} \gets \emptyset$
\ENDLOOP
\STATE return $K$
\end{algorithmic}
\end{algorithm}

To illustrate how the semi-na\"ive evaluation proceeds, consider the spanning tree example in Fig.~\ref{fig:running-example}. The CFG graph can be expressed as the following set of input tuples:
\begin{align*}
    \{&edge(l_1, l_2),\ edge(l_2, l_3),\ edge(l_3, l_4),\ edge(l_3, l_6), \\
    &edge(l_4, l_8),\ edge(l_6, l_8),\ edge(l_8, l_2),\ edge(l_2, l_{10})\}
\end{align*}

Then, to compute the spanning tree, we use the Datalog program in Fig.~\ref{fig:worklist-souffle}, consisting of two Datalog rules:
\begin{verbatim}
 .decl edge(v:symbol, u:symbol)
 .decl st(v:symbol, u:symbol) choice-domain u

 st("root","L1"). 
 st(v,u) :-  st(_, u),  edge(v,u).
\end{verbatim}

In the initial iteration, $I_0$, the spanning tree contains only the root node. In the following three iterations, the evaluation proceeds to add edges following the CFG:
\begin{align*}
    I_0 &= \{st(\texttt{root}, l_1)\} \\
    I_1 &= I_0 \cup \{st(l_1, l_2)\} \\
    I_2 &= I_1 \cup \{st(l_2, l_3), st(l_2, l_{10})\} \\
    I_3 &= I_2 \cup \{st(l_3, l_4), st(l_3, l_6)\} \\
\end{align*}

In iteration $I_4$, the choice constraint becomes important. The Datalog rules would compute both $st(l_4, l_8)$ and $st(l_6,l_8)$. However, keeping both tuples would violate the choice constraint on the second element. Therefore, the choice function arbitrarily chooses one of the tuples to keep, say $st(l_4, l_8)$:

\begin{align*}
    I_4 &= I_3 \cup \{st(l_4, l_8)\}
\end{align*}

In iteration $I_5$, the Datalog rules would compute $st(l_8, l_2)$. However, the tuple $st(l_1, l_2)$ already exists in the instance, and thus inserting $st(l_8, l_2)$ would violate the choice constraint. In this case, the exclusion function prevents the insertion of $st(l_8, l_2)$, and so we have
\begin{align*}
    I_5 = I_4 = \{&st(\texttt{root}, l_1),\ st(l_1, l_2),\ st(l_2, l_3),\ st(l_2, l_{10}), \\
                  &st(l_3, l_4),\ st(l_3, l_6),\ st(l_4, l_8)\}
\end{align*}
and a fixpoint is reached.

\newpage

\end{document}

%% file: souffle-pipeline.tex
\makeatletter
\pgfdeclareshape{document}{
\inheritsavedanchors[from=rectangle] 
\inheritanchorborder[from=rectangle]
\inheritanchor[from=rectangle]{center}
\inheritanchor[from=rectangle]{north}
\inheritanchor[from=rectangle]{south}
\inheritanchor[from=rectangle]{west}
\inheritanchor[from=rectangle]{east}
\backgroundpath{
\southwest \pgf@xa=\pgf@x \pgf@ya=\pgf@y
\northeast \pgf@xb=\pgf@x \pgf@yb=\pgf@y
\pgf@xc=\pgf@xb \advance\pgf@xc by-10pt 
\pgf@yc=\pgf@yb \advance\pgf@yc by-10pt
\pgfpathmoveto{\pgfpoint{\pgf@xa}{\pgf@ya}}
\pgfpathlineto{\pgfpoint{\pgf@xa}{\pgf@yb}}
\pgfpathlineto{\pgfpoint{\pgf@xc}{\pgf@yb}}
\pgfpathlineto{\pgfpoint{\pgf@xb}{\pgf@yc}}
\pgfpathlineto{\pgfpoint{\pgf@xb}{\pgf@ya}}
\pgfpathclose
\pgfpathmoveto{\pgfpoint{\pgf@xc}{\pgf@yb}}
\pgfpathlineto{\pgfpoint{\pgf@xc}{\pgf@yc}}
\pgfpathlineto{\pgfpoint{\pgf@xb}{\pgf@yc}}
\pgfpathlineto{\pgfpoint{\pgf@xc}{\pgf@yc}}
}
}
\makeatother

\tikzstyle{doc}=[%
draw,
thick,
align=center,
color=black,
shape=document,
minimum width=8mm,
minimum height=12mm,
shape=document,
inner sep=2ex,
font=\small,
]
\tikzstyle{arrow} = [thick,->,>=stealth]
\tikzstyle{process} = [rectangle, minimum width=2cm,text centered, draw=black,
fill=gray!30, text width=2.3cm, thick, font=\bfseries]
\begin{tikzpicture}[node distance=2cm]
\node [doc] (DL) {DL};

\node [right=0.5cm of DL] (parser) {
    \begin{tikzpicture}
      \node [process, rotate=90, text width=3cm] {Parser};    
    \end{tikzpicture}
};

\node [doc, right=1cm of parser] (AST) {AST};

\node [process, below=0.5cm of AST] (hl-opt) {High-Level\\Optimizations};

\node [above=0.5cm of AST] {Declarative};

\node [right=1cm of AST] (RAM-Trans) {
    \begin{tikzpicture}
      \node [process, rotate=90, text width=3.3cm] {RAM Transformer};
    \end{tikzpicture}
};

\node [doc, right=1cm of RAM-Trans] (RAM) {RAM};

\node [above=0.5cm of RAM] {Imperative};

\node [process, below=0.5cm of RAM] (ml-opt) {Mid-Level\\Optimizations};

\node [right=1cm of RAM] (Code-Gen) {
    \begin{tikzpicture}
      \node [process, rotate=90, text width=3cm ] {Code Generator};
    \end{tikzpicture}
};

\node [doc, right=1cm of Code-Gen] (CPP) {CPP};

\node [above=0.5cm of CPP] {Native};

\node [process, below=0.5cm of CPP] (Relation-Primitives) {Relational\\Primitives};

\node [right=1cm of CPP] (CPP-Compiler) {
    \begin{tikzpicture}
      \node [process, rotate=90, text width=3cm] {C++ Compiler};
    \end{tikzpicture}
};

\node [doc, right=0.5cm of CPP-Compiler] (BIN) {BIN};

\begin{scope}[on background layer]
    \draw [arrow] (DL) -- (AST);
    \draw [arrow] (AST) -- (RAM);
    \draw [arrow] (RAM) -- (CPP);
    \draw [arrow] (CPP) -- (BIN);
    \coordinate [below =0.03cm of AST] (ASTx);
    \coordinate [left =0.15cm of ASTx] (ASTxx);
    \coordinate [below =0.48cm of AST] (ASTy);
    \coordinate [right =0.15cm of ASTy] (ASTyy);
    \draw[arrow] (ASTxx) arc[radius=0.5, start angle=150, end angle=210] ;
    \draw[arrow] (ASTyy) arc[radius=-0.5, start angle=150, end angle=210] ;

    \coordinate [below =0.03cm of RAM] (RAMx);
    \coordinate [left =0.15cm of RAMx] (RAMxx);
    \coordinate [below =0.48cm of RAM] (RAMy);
    \coordinate [right =0.15cm of RAMy] (RAMyy);
    \draw[arrow] (RAMxx) arc[radius=0.5, start angle=150, end angle=210] ;
    \draw[arrow] (RAMyy) arc[radius=-0.5, start angle=150, end angle=210] ;

    \draw[arrow] (CPP) -- node[midway, right] {uses} (Relation-Primitives);
\end{scope}

\end{tikzpicture}